\documentclass[journal,10pt]{IEEEtran}
\usepackage{multicol}
\usepackage{multirow}
\usepackage{array}
\usepackage{amsmath}
\usepackage{algorithm}
\usepackage{algpseudocode}
\usepackage{amssymb}
\usepackage{amsfonts}
\usepackage{makecell}
\usepackage{textcomp}
\usepackage{xcolor}
\usepackage{stfloats}
\usepackage{url}
\usepackage{verbatim}
\usepackage{graphicx}
\usepackage{caption}
\usepackage{subcaption}
\usepackage{booktabs}
\def\BibTeX{{\rm B\kern-.05em{\sc i\kern-.025em b}\kern-.08em
    T\kern-.1667em\lower.7ex\hbox{E}\kern-.125emX}}
\usepackage{balance}

\begin{document}

\title{Intent-Driven 6G Communication Framework for RIS and Spectrum Leasing }

\author{
Zawar Hussain,~\IEEEmembership{Graduate Student Member,~IEEE}, 
Naveed Ul Hassan,~\IEEEmembership{Senior Member,~IEEE}
\thanks{
Zawar Hussain and Naveed Ul Hassan are with the Department of Electrical Engineering, 
Lahore University of Management Sciences (LUMS), Lahore, Pakistan 
(e-mail: zawar.hussain@lums.edu.pk; naveed.hassan@lums.edu.pk).
This research was supported by LUMS, Faculty Initiative Fund (FIF).
}
}

\maketitle

\begin{abstract}
Intent-Driven Communication (IDC) is emerging as a key paradigm for autonomous 6G networks, where AI and Large Language Models (LLMs) translate high-level user intents into actionable network policies. Meanwhile, Reconfigurable Intelligent Surfaces (RIS) and dynamic spectrum leasing are becoming essential for improving coverage and capacity in resource-constrained environments. This paper extends the IDC framework by integrating RIS and spectrum leasing into AI-assisted intent translation, policy mapping, and orchestration. A leasing-aware architecture is presented, and a Lyapunov-based Decision Support Framework is implemented as an illustrative mechanism for intelligent resource acquisition under time-varying prices and availability. Simulation results validate that the DSF achieves cost-efficient, delay-aware orchestration while exhibiting the expected Lyapunov stability properties. These findings highlight the feasibility of combining IDC with intelligent resource leasing in future 6G systems.

\end{abstract}

\begin{IEEEkeywords}
Intent-Driven Communication (IDC), Reconfigurable Intelligent Surface (RIS), Decision Support Framework (DSF), Lyapunov Optimization.
\end{IEEEkeywords}

\section{Introduction}

6G communication networks, as envisioned in the IMT-2030 framework, aim to support autonomous, service-aware, and resource-efficient connectivity for diverse applications, including industrial IoT, real-time control, and smart cities \cite{sharma2024role}. With increasing network complexity, Intent-Driven Communication (IDC) has emerged as a transformative paradigm that allows applications to specify high-level requirements, e.g., latency, reliability, or cost constraints, while the network autonomously determines how to fulfill these requirements \cite{Njah2025An}. Recent advances in Artificial Intelligence (AI) and Large Language Models (LLMs) further enhance IDC by enabling natural-language intent interpretation, policy reasoning, and closed-loop assurance \cite{11112781}. This enables autonomous translation of high-level user intents into actionable, low-level executable configurations, supporting intelligent and adaptive 6G systems \cite{IDN_Survey}. 

In LLM-enabled IDC architectures, LLMs operate primarily in the intent and management plane, translating high-level user requirements into structured, machine-interpretable policies. Recent studies on LLM-assisted wireless systems position LLMs as reasoning and decision-support engines for network optimization and resource management rather than physical-layer signal processing mechanisms \cite{LLM_Wireless_Survey,LLM_Resource_Allocation}. This architectural separation aligns with the OSI model, where intent processing resides above the traditional protocol stack, while scheduling and transmission control are executed at lower layers.

\begin{figure*}[t]
\centering
\includegraphics[width=0.99\linewidth, height=0.70\linewidth]
{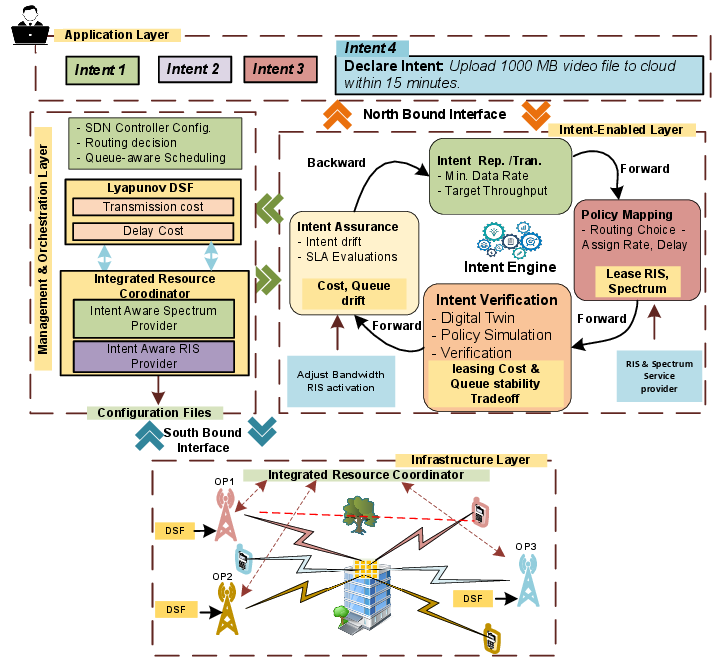}
\caption{Schematic of the IDC framework, illustrating the four layers and their interactions in transforming high-level user intents into network-level actions in a 6G environment.}
\label{Fig:echo system}
\end{figure*}

The proposed IDC framework, illustrated in Fig.~\ref{Fig:echo system}, comprises four layers: (i) Application Layer, where service intents are declared, (ii) Intent-Enabled Layer, which performs intent translation, policy mapping, and assurance, (iii) Management and Orchestration (M\&O) Layer, which converts policies into device configurations, and (iv) Infrastructure Layer, which executes the resulting actions on network elements. This architecture enables seamless integration of emerging technologies such as RIS and dynamic spectrum leasing, allowing structured transformation of user intent into network-level actions. \cite{basharat2021reconfigurable,Spectrum}. Prior studies have shown that intelligent spectrum leasing combined with Lyapunov optimization can effectively balance leasing cost and QoS requirements in dynamic wireless environments \cite{hassan2017exploiting, hassan2017guaranteeing}. Motivated by this synergy, this paper integrates RIS and spectrum leasing into an intent-driven resource orchestration framework. Extending prior Lyapunov-based approaches for cost-aware spectrum leasing and sharing \cite{joshi2017dynamic, hussain2025ris}, we develop a fully intent-driven architecture in which a Lyapunov-based Decision Support Framework (DSF) dynamically minimizes leasing cost while satisfying user QoS requirements through a threshold-based policy that adapts to real-time system states and price variations \cite{hussain2025joint}. By embedding RIS and spectrum leasing entities across intent translation, policy mapping, verification, and assurance stages, the proposed framework enables autonomous, cost-efficient, and QoS-compliant orchestration of external resources to fulfill high-level user intents in emerging 6G networks.

Although the IDC framework spans multiple architectural layers to illustrate end-to-end intent flow, the contributions of this work go beyond introducing the LLM-enabled IDC framework. In particular, this work also develops an intent-aware MAC-layer resource scheduling mechanism under dynamic leasing. The proposed Lyapunov-based DSF operates at the MAC layer, which performs slot-level, queue-aware scheduling and dynamic RIS/spectrum leasing decisions under time-varying traffic and price conditions. Physical-layer RIS operations, such as phase-shift configuration and beamforming, are abstracted and assumed to be handled by the RIS controller once leasing is activated.

To illustrate a practical application of this framework, we consider a delay-constrained video upload scenario in which user intent is translated into QoS requirements and enforced through the proposed Lyapunov-based MAC-layer control mechanism. This illustrates how LLM-enabled intent processing can be systematically integrated with queue-aware resource scheduling for cost-efficient and delay-aware orchestration in emerging 6G networks.

\begin{figure*}[t]
\centering
\includegraphics[width=0.9\linewidth, height=0.68\linewidth]
{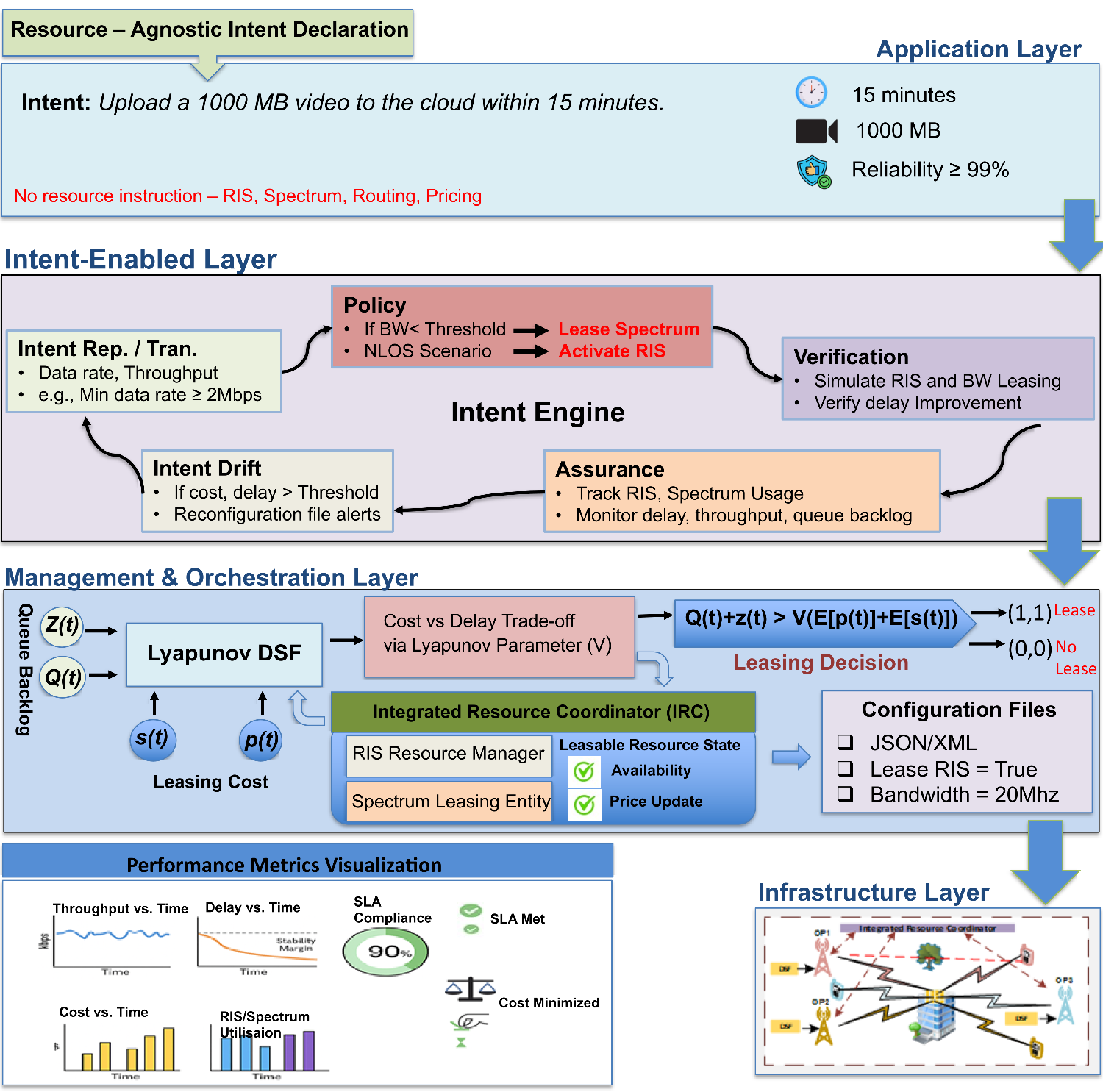}
\vspace{2.0mm}
\caption{Illustrative example of the Lyapunov-based DSF applied to a video upload scenario. The figure shows how queue dynamics and time-varying leasing prices guide RIS and spectrum acquisition decisions, demonstrating cost- and delay-aware intent fulfillment within the IDC framework.}

\label{System Model}
\end{figure*}

The rest of the paper is organized as follows. We present the IDC framework-based system model in Section II, followed by problem formulation and decision support framework in Section III. Section IV contains simulation results, and we conclude the paper in Section V.

\section{Leasing-Aware IDC System Model: Video Transmission Use Case}

We consider a leasing-aware IDC framework operating in a discrete-time 6G uplink scenario as illustrated in Fig.~\ref{System Model}. In this scenario, a network must fulfill a high-level user intent generated by a smart city operator to upload a video. The data is collected via smart devices and transmitted by dynamically acquiring external RIS and spectrum resources. The framework translates resource-agnostic user intents into network-level actions through a layered architecture. 

The process begins at the \textit{Application Layer}, where the user declares a high-level intent such as \textit{``Upload a 1000 MB video within 15 minutes, ensuring at least 99\% reliability''}. This request is forwarded to the \textit{Intent-Enabled Layer}, where the intent engine translates the user requirement into network parameters such as minimum data rate and delay constraints. These parameters are mapped to resource policies specifying when RIS or additional spectrum resources should be leased. A digital-twin-based verification module evaluates policy feasibility, while assurance mechanisms monitor metrics such as delay, throughput, spectrum usage, and queue backlog to detect intent drift.

The translated leasing-aware policies are executed in the \textit{M\&O Layer},  which functions as the policy execution engine. This layer generates device-specific configuration files through a southbound interface based on the optimal policies derived by the Intent-Enabled Layer. A Lyapunov drift-plus-penalty DSF dynamically determines leasing actions based on queue backlog and time-varying resource prices. An Integrated Resource Coordinator (IRC) interfaces with the RIS Resource Manager (RRM) and Spectrum Leasing Entity (SLE) to obtain real-time availability and pricing information. The resulting leasing decisions are converted into configuration files and dispatched through the southbound interface to infrastructure layer controllers.

Finally, the \textit{Infrastructure Layer} performs data transmission and resource activation using base stations, RIS nodes, and SDN-enabled switches. These components also provide telemetry including queue length, link quality, and resource utilization, enabling closed-loop intent assurance. 
RIS phase-shift optimization to improve signal quality and spectral efficiency has been extensively studied \cite{9140329}. Our focus is on intent-driven orchestration and leasing decisions at the M\&O layer rather than on physical-layer beamforming optimization. RIS phase configuration is handled by the intent-aware RIS provider under the Integrated Resource Coordinator (IRC), which performs local channel estimation and phase adjustment to maximize link performance. Thus, RIS phase optimization is treated as an internal functionality of the leased RIS resource, while the proposed framework determines when to lease RIS resources based on system-level objectives.
This system model illustrates how high-level user intents are translated into resource-aware actions in a 6G network, enabling intelligent leasing decisions within the IDC framework.

\section{Lyapunov-Based Decision Support Framework Example}

To demonstrate the practical realization of intent-driven resource orchestration within the proposed IDC framework, we illustrate an example implementation using a Lyapunov-based DSF at the M\&O layer. The DSF performs slot-level leasing decisions based on queue congestion and time-varying leasing prices. 

We consider a discrete-time system where data units arrive randomly and are queued before transmission. The network maintains a data queue \( Q(t) \)  and virtual queue \( Z(t) \) to handle queue delay and penalize non-transmission, respectively. The queues evolve as follows:
\begin{equation}
    Q(t+1) = \max(Q(t) - R(t), 0) + A(t+1),
\end{equation}
\begin{equation}
Z(t+1) = \max\big(Z(t) - R(t) + \epsilon_d \cdot [1 - R(t)], 0\big)
\end{equation}
Here, \(Q(t)\) represents the number of discretized data units (e.g., packets) awaiting transmission at the beginning of time slot \(t\). The term \(\max(Q(t)-R(t),0)\) captures the remaining backlog after transmission during slot \(t\), where the max operator ensures that the queue length remains non-negative. The variable \(A(t+1)\) denotes the arrival of a new data unit in the subsequent time slot. The virtual queue \(Z(t)\) captures delay urgency by accumulating penalties when transmission is deferred. Specifically, when no transmission occurs (\(R(t)=0\)), the penalty term \(\epsilon_d\) increases \(Z(t)\), enforcing delay awareness in the leasing decision process and discouraging prolonged queue buildup. This formulation follows standard Lyapunov queuing models with binary arrivals \(A(t)\) and departures \(R(t)\). Successful transmission requires both RIS and spectrum resources, so the 
departure process satisfies:
\begin{equation}
    R(t) = x(t) \, y(t),
\end{equation}
where \( x(t), y(t) \in \{0,1\} \) represent the binary leasing decisions for 
RIS and spectrum, respectively. The instantaneous transmission cost is:
\begin{equation}
    C(t) = x(t) p(t) + y(t) s(t),
\end{equation}
where \( p(t) \) and \( s(t) \) denote the time-varying leasing prices of leasing RIS and spectrum. The objective is to minimize the time-average expected cost of transmission while maintaining a stable data queue. The Lyapunov function measures the overall system congestion and is defined as:
\begin{equation}
    \mathcal{L}(Q(t), Z(t)) = \frac{1}{2} \big( Q^2(t) + Z^2(t) \big),
\end{equation}
Using \( \Theta(t) = (Q(t), 
Z(t)) \), we define the drift-plus-penalty method. The DSF minimizes an upper bound on:
\begin{equation}
    \Delta(\Theta(t)) + V \, \mathbb{E}[ C(t) \mid \Theta(t) ],
\end{equation}
Simplifying the above yields the per-slot objective function \( G(t) \), which we need to minimize in each time slot given by:  
\begin{equation}
\begin{split}
G(t) = V \big[x(t)\cdot \mathbb{E}[p(t)]+y(t)\cdot \mathbb{E} [s(t)] \big] - Q(t) \cdot R(t)\\
- Z(t) \cdot \big[ \epsilon_d \cdot (1 - R(t))\big]
\end{split}
\end{equation}
where \( V > 0 \) is the Lyapunov weight parameter that governs the trade-off between cost minimization and queue stability, and \( \epsilon_d > 0 \) is the delay penalty parameter that regulates the growth of the virtual queue \( Z(t) \) when transmission is deferred.
The resulting threshold-based decision rule is:

\begin{equation}
(x(t), y(t)) =
\begin{cases}
(1,1), &
\text{if } Q(t) + Z(t) > V\big(\mathbb{E}[p(t)]
\\[-1pt]
& \qquad\qquad\qquad\quad + \mathbb{E}[s(t)]\big)
\\[6pt]
(0,0), & \text{otherwise}.
\end{cases}
\end{equation}

The DSF ensures that leasing occurs only when congestion exceeds a cost-weighted threshold. In practical intent-driven deployments, the control parameters $(V, \epsilon_d)$ can be initialized based on high-level service requirements before DSF execution. Larger $\epsilon_d$ values are assigned to delay-critical services to accelerate virtual queue growth, while larger $V$ values prioritize cost-sensitive operation by raising the leasing threshold. Pre-trained LLMs at the Intent-Enabled Layer can assist this initialization via prompt-based intent interpretation without task-specific retraining or joint optimization \cite{LLM_Telecom_Survey}. Once configured, the Lyapunov-based DSF operates fully online using fixed parameter values and real-time queue and price observations. No joint training or online optimization of $(V, \epsilon_d)$ with the LLM is required.

The resulting leasing decisions are communicated to the IRC, which interfaces with the RIS resource manager and spectrum leasing entities to verify resource availability, retrieve updated prices, and execute leasing requests. These leasing actions are then translated into device-specific configuration files (JSON/XML) and applied to SDN controllers, base stations, RIS nodes, and spectrum controllers via the southbound interface. In the Infrastructure Layer, the leased RIS and spectrum resources enhance uplink propagation, enabling the successful transmission of queued packets. Real-time telemetry, including queue length, SNR, RIS activation status, and spectrum usage, is continuously fed back to the Intent-Enabled Layer to support service-level agreement evaluation and closed-loop intent assurance. This example illustrates how the IDC framework can operationalize high-level intents through dynamic, cost-aware resource orchestration while maintaining alignment with user requirements.

The computational overhead associated with AI-assisted intent interpretation operates at a slower control-plane timescale than the slot-level operation of the Lyapunov-based DSF and is therefore not explicitly modeled. Recent LLM-enabled 6G studies indicate that edge-based LLM inference incurs non-negligible latency and resource consumption, advocating hierarchical architectures in which LLMs perform high-level orchestration rather than real-time PHY/MAC scheduling~\cite{wang2025edge}. In the proposed framework, the LLM is invoked only for intent translation and parameter initialization. Once configured, the Lyapunov-based DSF executes lightweight threshold-based decisions online with negligible per-slot complexity. Hence, LLM inference delay does not lie in the real-time control loop. Modeling computation-aware AI latency is left for future work.

\section{Simulation Results}

To evaluate the effectiveness of the proposed leasing-aware IDC framework, we simulate an uplink video upload scenario in which RIS and spectrum availability evolve as independent binary processes and leasing prices vary uniformly over $[1,10]$ cents per time slot. These simulations illustrate how high-level user intents are operationalized through the IDC pipeline, with AI and LLMs assisting in intent translation and policy mapping, and the Lyapunov-based Decision Support Framework (DSF) serving as an illustrative mechanism for dynamic leasing decisions. The goal is to demonstrate that integrating intelligent resource leasing within a closed-loop IDC system enables both cost and delay-aware orchestration, all while ensuring alignment with the original user intent.

\begin{figure}[t]
\centerline{\includegraphics[width=9cm, height=7cm]
{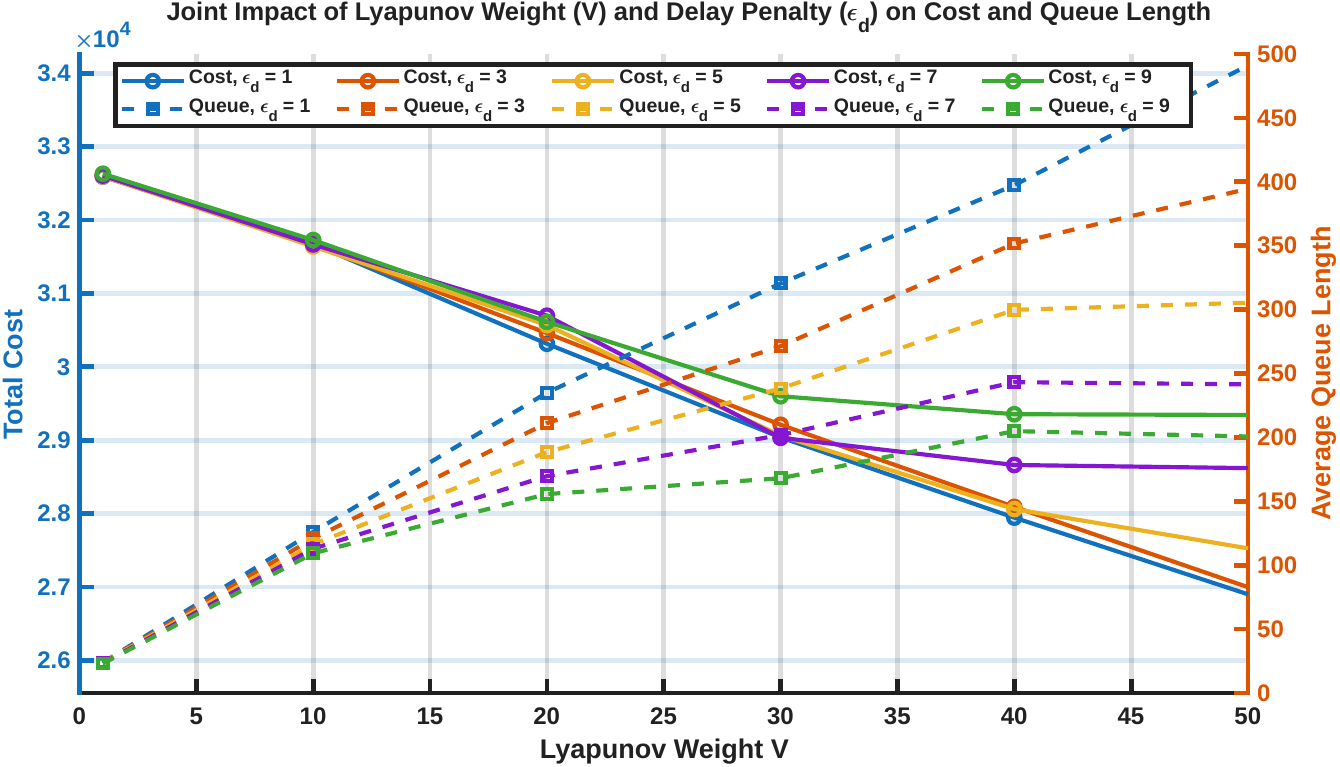}}
\caption{Cost and queue length for $T=5000$ time slots}
\label{fig_cost_queue}
\end{figure}

\begin{figure}[t]
\centerline{\includegraphics[width=9cm, height=7.0cm]{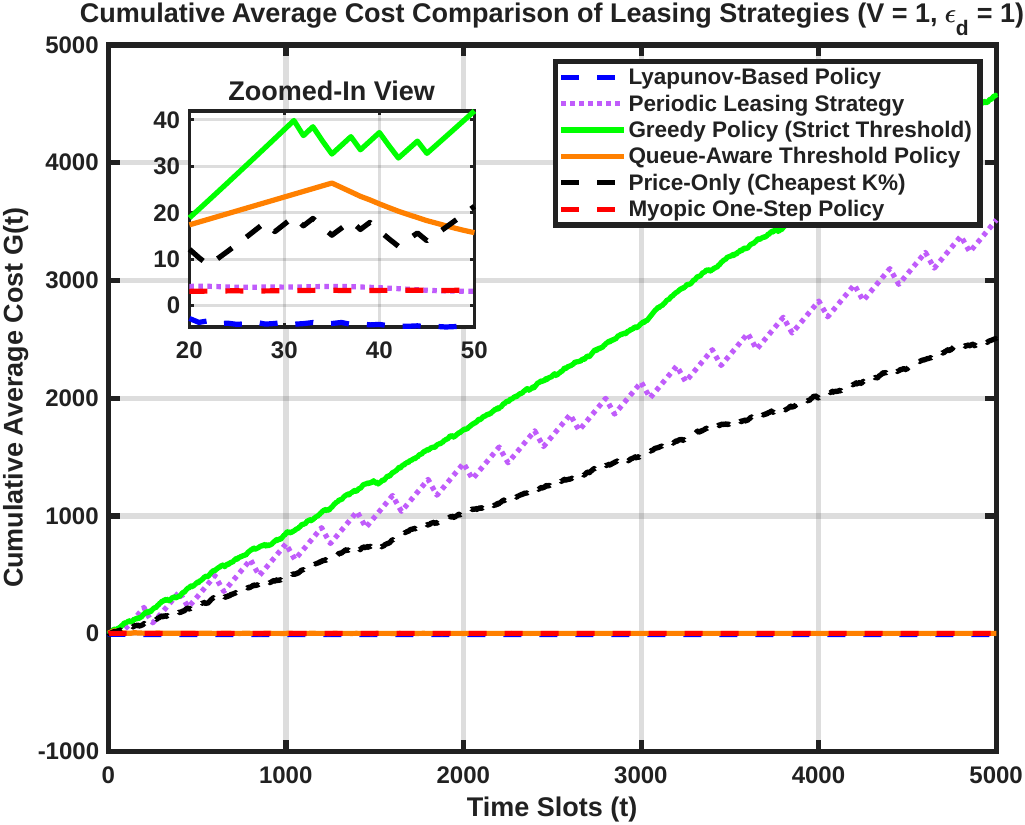}}
\caption{Cumulative average cost comparison with baseline strategies.}
\label{fig_baseline_hybrdi}
\end{figure}

Fig.~\ref{fig_cost_queue} shows the joint impact of the Lyapunov weight $V$ and delay penalty $\epsilon_d$ on accumulated leasing cost and average queue length. Increasing $V$ prioritizes cost efficiency, resulting in fewer leasing actions but larger queue backlogs. Conversely, increasing $\epsilon_d$ strengthens delay awareness, prompting more aggressive leasing decisions that maintain shorter queues at the expense of higher cost. These results illustrate how the IDC framework can balance competing objectives through adaptive resource orchestration. Fig.~\ref{fig_baseline_hybrdi} compares the cumulative average cost $G(t)$ of the proposed Lyapunov-based policy with several benchmark strategies, including periodic, greedy, price-only, queue-threshold, and myopic policies. Unlike these approaches, the proposed DSF jointly considers queue stability and leasing cost through drift-plus-penalty control, achieving lower cumulative cost while maintaining stable queue behavior.

Although the DSF serves as an illustrative mechanism, AI and LLMs underpin the interpretation of high-level user intents and guide policy mapping, ensuring that the resource orchestration aligns with the desired service-level objectives. These simulations demonstrate that the IDC framework can operationalize intents through intelligent, adaptive leasing while balancing cost and delay considerations in scenarios representative of real-world 6G networks.

\section{Conclusion}

This paper presents a leasing-aware IDC framework for 6G networks that integrates RIS and dynamic spectrum acquisition mechanism to support intent-driven services. Lyapunov-based DSF is demonstrated as an example mechanism to balance cost efficiency and delay performance using a threshold-based leasing rule that accounts for queue dynamics and time-varying prices. Simulation results demonstrate that the proposed approach achieves cost-efficient and delay-aware orchestration while maintaining queue stability and outperforming several baseline strategies. The results highlight the potential of combining IDC with intelligent resource leasing for future autonomous 6G networks. 

\bibliographystyle{IEEEtran}   
\bibliography{references}

@ARTICLE{IDN_Survey,
  author={Wang, Yao and Yang, Chungang and Li, Tong and Ouyang, Ying and Mi, Xinru and Song, Yanbo},
  journal={IEEE Communications Surveys \& Tutorials}, 
  title={A Survey on Intent-Driven End-to-End 6G Mobile Communication System}, 
  year={2025},
  volume={},
  number={},
  pages={1-1},
  doi={10.1109/COMST.2025.3575041}}

@article{basharat2021reconfigurable,
  title={Reconfigurable intelligent surfaces: Potentials, applications, and challenges for {6G} wireless networks},
  author={Basharat, Sarah and Hassan, Syed Ali and Pervaiz, Haris and Mahmood, Aamir and Ding, Zhiguo and Gidlund, Mikael},
  journal={IEEE Wireless Communications},
  volume={28},
  number={6},
  pages={184--191},
  year={2021},
  publisher={IEEE}
}

@ARTICLE{Spectrum,
  author={Na, Minsoo and Lee, Jaehyun and Choi, Giwan and Yu, Takki and Choi, Jeongsik and Lee, Jinyoung and Bahk, Saewoong},
  journal={IEEE Communications Magazine}, 
  title={Operator's Perspective on 6G: 6G Services, Vision, and Spectrum}, 
  year={2024},
  volume={62},
  number={8},
  pages={178-184},
  doi={10.1109/MCOM.001.2400060}}

@article{joshi2017dynamic,
  title={Dynamic inter-operator spectrum sharing via Lyapunov optimization},
  author={Joshi, Satya Krishna and Manosha, KB Shashika and Codreanu, Marian and Latva-aho, Matti},
  journal={IEEE Transactions on Wireless Communications},
  volume={16},
  number={10},
  pages={6365--6381},
  year={2017},
  publisher={IEEE}
}

@INPROCEEDINGS{hussain2025ris,
  author={Hussain, Zawar and Hassan, Naveed Ul and Khan, Ali Hussain and Naqvi, Ijaz},
  booktitle={2025 2nd International Conference on Microwave, Antennas \& Circuits (ICMAC)}, 
  title={Cost Optimization for Reconfigurable Intelligent Surface Resource Leasing in 6G Networks}, 
  year={2025},
  volume={},
  number={},
  pages={1-4},
  doi={10.1109/ICMAC64768.2025.11003270}}

@inproceedings{hussain2025joint,
  title={Joint RIS Leasing and Energy Management for Delay-Tolerant 6G Networks},
  author={Hussain, Zawar and Hassan, Naveed Ul and Naqvi, Ijaz Haider},
  booktitle={2025 7th Global Power, Energy and Communication Conference (GPECOM)},
  pages={1067--1072},
  year={2025},
  organization={IEEE}
}

@article{Njah2025An,
title={An AI-Driven Intent-Based Network Architecture},
author={Yosra Njah and Aris Leivadeas and M. Falkner},
journal={IEEE Communications Magazine},
year={2025},
volume={63},
pages={146-153},
doi={10.1109/mcom.001.2400143}
}

@article{sharma2024role,
  title={The role of 6G technologies in advancing smart city applications: Opportunities and challenges},
  author={Sharma, Sanjeev and Popli, Renu and Singh, Sajjan and Chhabra, Gunjan and Saini, Gurpreet Singh and Singh, Maninder and Sandhu, Archana and Sharma, Ashutosh and Kumar, Rajeev},
  journal={Sustainability},
  volume={16},
  number={16},
  pages={7039},
  year={2024},
  publisher={MDPI}
}

@ARTICLE{11112781,
  author={Habib, Md Arafat and Iturria-Rivera, Pedro Enrique and Ozcan, Yigit and Elsayed, Medhat and Bavand, Majid and Gaigalas, Raimundas and Erol-Kantarci, Melike},
  journal={IEEE Transactions on Network Science and Engineering}, 
  title={Harnessing the Power of LLMs, Informers and Decision Transformers for Intent-Driven RAN Management in 6G}, 
  year={2025},
  volume={},
  number={},
  pages={1-20},
  doi={10.1109/TNSE.2025.3596028}}

@ARTICLE{LLM_Wireless_Survey,
  author={Zhou, Hao and Hu, Chengming and Yuan, Dun and Yuan, Ye and Wu, Di and Chen, Xi and Tabassum, Hina and Liu, Xue},
  journal={IEEE Wireless Communications}, 
  title={Large Language Models for Wireless Networks: An Overview from the Prompt Engineering Perspective}, 
  year={2025},
  volume={32},
  number={4},
  pages={98-106},
  doi={10.1109/MWC.001.2400384}}

@ARTICLE{LLM_Resource_Allocation,
  author={Lee, Woongsup and Park, Jeonghun},
  journal={IEEE Access}, 
  title={LLM-Empowered Resource Allocation in Wireless Communications Systems}, 
  year={2026},
  volume={14},
  number={},
  pages={15260-15272},
  doi={10.1109/ACCESS.2026.3655801}}

@ARTICLE{LLM_Telecom_Survey,
  author={Zhou, Hao and Hu, Chengming and Yuan, Ye and Cui, Yufei and Jin, Yili and Chen, Can and Wu, Haolun and Yuan, Dun and Jiang, Li and Wu, Di and Liu, Xue and Zhang, Jianzhong and Wang, Xianbin and Liu, Jiangchuan},
  journal={IEEE Communications Surveys \& Tutorials}, 
  title={Large Language Model (LLM) for Telecommunications: A Comprehensive Survey on Principles, Key Techniques, and Opportunities}, 
  year={2025},
  volume={27},
  number={3},
  pages={1955-2005},
  doi={10.1109/COMST.2024.3465447}}

@article{wang2025edge,
  title={Edge large AI models: Collaborative deployment and IoT applications},
  author={Wang, Zixin and Shi, Yuanming and Letaief, Khaled B},
  journal={IEEE Internet of Things Magazine},
  year={2025},
  publisher={IEEE}
}

@ARTICLE{9140329,
  author={Di Renzo, Marco and Zappone, Alessio and Debbah, Merouane and Alouini, Mohamed-Slim and Yuen, Chau and de Rosny, Julien and Tretyakov, Sergei},
  journal={IEEE Journal on Selected Areas in Communications}, 
  title={Smart Radio Environments Empowered by Reconfigurable Intelligent Surfaces: How It Works, State of Research, and The Road Ahead}, 
  year={2020},
  volume={38},
  number={11},
  pages={2450-2525},
  doi={10.1109/JSAC.2020.3007211}}

@inproceedings{hassan2017exploiting,
  title={Exploiting QoS flexibility for smart grid and IoT applications using TV white spaces},
  author={Hassan, Naveed Ul and Yuen, Chau and Atique, Muhammad Bershgal},
  booktitle={2017 IEEE International Conference on Communications (ICC)},
  pages={1--6},
  year={2017},
  organization={IEEE}
}

@article{hassan2017guaranteeing,
  title={Guaranteeing QoS using unlicensed TV white spaces for smart grid applications},
  author={Hassan, Naveed Ul and Tushar, Wayes and Yuen, Chau and Kerk, See Gim and Oh, Ser Wah},
  journal={IEEE Wireless Communications},
  volume={24},
  number={2},
  pages={18--25},
  year={2017},
  publisher={IEEE}
}

\end{document}